\documentclass[superscriptaddress,aps,preprintnumbers,amsmath,showpacs,amssymb,nofootinbib,reprint]{revtex4-1}
\usepackage{bm, color} 
\usepackage{amssymb,amsfonts,slashed,amsthm,amsmath,graphicx, soul}
\usepackage{subfigure}
\usepackage{hyperref}
\usepackage{tikz,xcolor}

\begin{document}

\title{Nambu-Goldstone emissions from the cosmological evolution of global monopoles}

\author{Yukihiro Kanda}
\affiliation{Institute for Cosmic Ray Research, The University of Tokyo, \\
Kashiwa, Chiba, 277-8582, Japan}

\author{Naoya Kitajima}
\affiliation{Frontier Research Institute for Interdisciplinary Sciences, Tohoku University, \\
6-3 Azaaoba, Aramaki, Aoba-ku, Sendai 980-8578, Japan}
\affiliation{Department of Physics, Tohoku University, \\
6-3 Azaaoba, Aramaki, Aoba-ku, Sendai 980-8578, Japan}


\begin{abstract}
We show the emission of the Nambu-Goldstone (NG) bosons from the cosmological evolution of global monopoles. The NG bosons are non-thermally produced from the dynamics of global monopoles such as the pair annihilation of monopole and anti-monopole. Our numerical lattice simulations demonstrate that the spectrum of NG bosons emitted from the scaling evolution of global monopoles has a peak around the horizon scale, as in the case with the global string and the semilocal string. We also estimate the abundance of the pseudo-Nambu-Goldstone (pNG) dark matter when one (or both) of the NG modes has a soft mass.
\end{abstract}

\maketitle

\section{Introduction}

Topological defects can be formed during cosmological phase transitions associated with spontaneous symmetry breaking in the early universe~\cite{Kibble:1976sj}. As possible relics of physics beyond the Standard Model, they may leave distinctive imprints on the subsequent evolution of the universe~\cite{Vilenkin:2000jqa}. Understanding the cosmological evolution of defect network and the observable signatures they produce is therefore important for connecting theories of high-energy physics beyond the Standard Model with cosmological observations. 

One important cosmological role of topological defects is that they can act as non-thermal sources of particles such as axion emissions from global strings \cite{Davis:1986xc,Vilenkin:1986ku,Garfinkle:1987yw,Yamaguchi:1998gx,Hagmann:2000ja,Hiramatsu:2010yu} and local strings \cite{Kitajima:2025jct}, and (light) dark photon emissions from local strings \cite{Long:2019lwl,Kitajima:2022lre,Kitajima:2023vre}. 
Since the axion emitted from the global string network can make a significant contribution to the present axion dark matter abundance in some cosmological history~\cite{Kawasaki:2014sqa,Gorghetto:2020qws}, an accurate prediction of the abundance requires a precise understanding of both the evolution of the string network and the spectrum of the emitted axions. In fact, some groups have reported a departure from the scaling behavior in which the number of long strings per Hubble volume remains constant~\cite{Gorghetto:2018myk,Kawasaki:2018bzv,Gorghetto:2020qws,Buschmann:2021sdq,Saikawa:2024bta,Benabou:2024msj}.\footnote{
Other groups have shown numerical results supporting the scaling law~\cite{Hindmarsh:2019csc,Hindmarsh:2021vih,Correia:2024cpk}, and thus the scaling violation of the global string network is still under discussion.
}
Such a departure can substantially modify the predicted axion dark matter abundance compared with the estimate based on the exact scaling. 

The global monopole is another kind of stable topological defects based on a global symmetry and it also has a channel to emit Nambu-Goldstone (NG) bosons~\cite{Vilenkin:2000jqa,Barriola:1989hx,Achucarro:2000td}. Due to the long-range interaction between global monopoles, monopole-antimonopole pair annihilation proceeds efficiently, and the network approaches the scaling regime~\cite{Bennett:1990xy,Turok:1991qq,Yamaguchi:2001rf,Yamaguchi:2001xn,Martins:2008ks,Lopez-Eiguren:2016jsy,Sousa:2017wvx}.\footnote{
The logarithmic deviation from the scaling evolution of the global monopole network is also reported in~\cite{Nakano:2026zme} like the global string case.
}
During its evolution, the monopole network is thought to lose most of its energy through the emission of NG bosons~\cite{Barriola:1989hx}. Nevertheless, the properties of this radiation, particularly its energy spectrum, have not yet been studied in detail. If the NG bosons subsequently acquire a mass, for example through a small explicit symmetry breaking of the underlying symmetry, they become pseudo-NG (pNG) bosons and may contribute to the present dark matter abundance. A quantitative prediction of this contribution requires not only the total amount of energy emitted from the network but also the evolution of the spectrum of the emitted NG bosons. 

In this paper, we investigate the NG boson emission from the evolution of the global monopole network using numerical lattice simulations, with particular emphasis on the spectrum of the radiated modes. We confirm that the monopole network reaches a scaling regime and efficiently radiates NG bosons during its evolution. We find that the emitted NG bosons have a soft spectrum with a characteristic energy of the order of the Hubble scale. By tracking the evolution of their number density, we quantify their contribution to the present dark matter abundance in scenarios where the NG bosons acquire a small mass and become pNG bosons. 

A further motivation of this study is to clarify the origin of NG boson emission from the network of semilocal strings~\cite{Vachaspati:1991dz,Hindmarsh:1991jq}. In our previous work, we found that NG bosons are efficiently produced during the evolution of a semilocal string network, as in the case of global string networks~\cite{Kanda:2025hgi}. However, unlike global strings, semilocal strings are not surrounded by a winding configuration of the NG modes. The dominant emission mechanism in the semilocal system must therefore differ from that of global strings, and identifying this remains an open question. 

Semilocal string segments are often described heuristically as local strings terminated by global-monopole-like endpoints \cite{Gibbons:1992gt,Hindmarsh:1992yy}, and this picture successfully captures several qualitative features of their network evolution \cite{Achucarro:2005vpt,Nunes:2011sf,Achucarro:2013mga,Lopez-Eiguren:2017ucu,Achucarro:2019blr}. It is therefore natural to expect that the dynamics and radiation properties of global monopoles may provide insight into the emission operating in semilocal string networks. 
In this paper, we show that the spectrum of the NG bosons emitted from the global monopole network closely resembles that obtained for the semilocal string network. This similarity provides strong evidence that the NG boson emission in the semilocal system is predominantly generated by the dynamics of the string endpoints.

This paper is organized as follows.
In Sec.~\ref{sec:model}, we introduce the model and show the extraction of NG modes. In Sec.~\ref{sec:numerical}, we show the numerical setup and results. In Sec.~\ref{sec:pNGDM}, we estimate the relic abundance of the emitted NG bosons after they acquire a mass and become pNG dark matter. Finally, we give discussions in Sec.~\ref{sec:discussion}.

\section{Global monopole and NG boson} \label{sec:model}

\subsection{Model}

In this paper, we consider the model based on the global $O(3)$ symmetry, containing three real scalar fields (scalar triplet) $\phi_i$ with $i=1,2,3$.
The Lagrangian of this model is
\begin{align}
    \mathcal{L} = \sum_{i=1}^3\frac{1}{2} \partial_\mu \phi_i \partial^\mu \phi_i - V(\phi_i),
\end{align}
where the scalar potential is given by
\begin{align}
V(\phi_i) = \frac{\lambda}{4} (\phi_1^2 + \phi_2^2 + \phi_3^2 - v^2)^2,
\end{align}
with $v$ being the vacuum expectation value and $\lambda$ the dimensionless self-coupling constant.

When the $O(3)$ symmetry is spontaneously broken, the radial component, $\phi_r \equiv \sqrt{\phi_1^2+\phi_2^2+\phi_3^2}$, acquires the vacuum expectation value, i.e. $\langle \phi_r \rangle = v$, and, accordingly, the field takes values on the $S_2$ vacuum manifold. Consequently, global monopoles can form. The field configuration of the global monopole is given by~\cite{Barriola:1989hx}
\begin{align}
    \phi_i = v f(r) \hat{x}^i ,
\end{align}
where $\hat{x}^i$ denotes the radial unit vector. The profile function $f(r)$ increases monotonically from $f(0)=0$ to $f(\infty)=1$. The characteristic monopole core radius is 
$(\sqrt{2\lambda}v)^{-1}$. 

Since the gradient energy of the Goldstone fields outside the monopole core dominates the monopole energy, the energy of a monopole-antimonopole pair separated by a distance $R$ can be estimated as $E\simeq 4\pi v^2 R$. The resulting attractive force is therefore approximately independent of their separation, with a magnitude $F \simeq 4\pi v^2$. A monopole-antimonopole pair accelerates toward each other and reaches relativistic velocities, so that its lifetime is of the order $R$. The energy lost by the pair during this process is expected to be carried away mainly in the form of NG boson radiation~\cite{Barriola:1989hx}.

To study the cosmological evolution of the global monopole network, we have to follow the fully nonlinear dynamics of the scalar field system in the expanding background.
The evolution equations for scalar fields in the flat Friedmann-Lema\^{i}tre-Robertson-Walker (FLRW) background spacetime are given as follows,
\begin{align} \label{eq:KGeq}
\ddot{\phi}_i + 3H \dot{\phi}_i - \frac{\nabla^2 \phi_i}{a^2} + \frac{\partial V}{\partial \phi_i} = 0,
\end{align}
where the overdot represents the derivative with respect to the cosmic time, $a$ is the scale factor, $H = \dot{a}/a$ is the Hubble parameter.
In what follows, we assume the radiation dominated universe, i.e., $H = 1/(2t)$.

\subsection{NG boson extraction}

After the breaking of the $O(3)$ symmetry, the system has one massive mode and two massless NG modes.
To extract the NG modes, let us express the scalar triplet as follows
\begin{align}
\phi_1 &= \phi_r \sin\vartheta \cos\varphi,\\
\phi_2 &= \phi_r \sin\vartheta \sin\varphi,\\
\phi_3 &= \phi_r \cos\vartheta,
\end{align}
where $\phi_r$ is the radial mode and $\vartheta$ and $\varphi$ are degrees of freedom on the vacuum manifold, corresponding to two massless NG modes.
Specifically, the kinetic energy of the NG modes are given by
\begin{align}
    \rho_K^{({\rm NG})} = \frac{1}{2} v^2 \dot\vartheta^2 + \frac{1}{2} v^2 \sin^2\vartheta \dot\varphi^2,
\end{align}
which is used to calculate the spectrum of emitted NG modes.
Here after we call the first and second NG modes (NG1 and NG2) for the first and second terms, respectively, on the right-hand side in the above equation.

\section{Numerical analysis} \label{sec:numerical}

\subsection{Setup}

\begin{figure*}[tp]
\centering
\includegraphics[width = 17.5cm, clip]{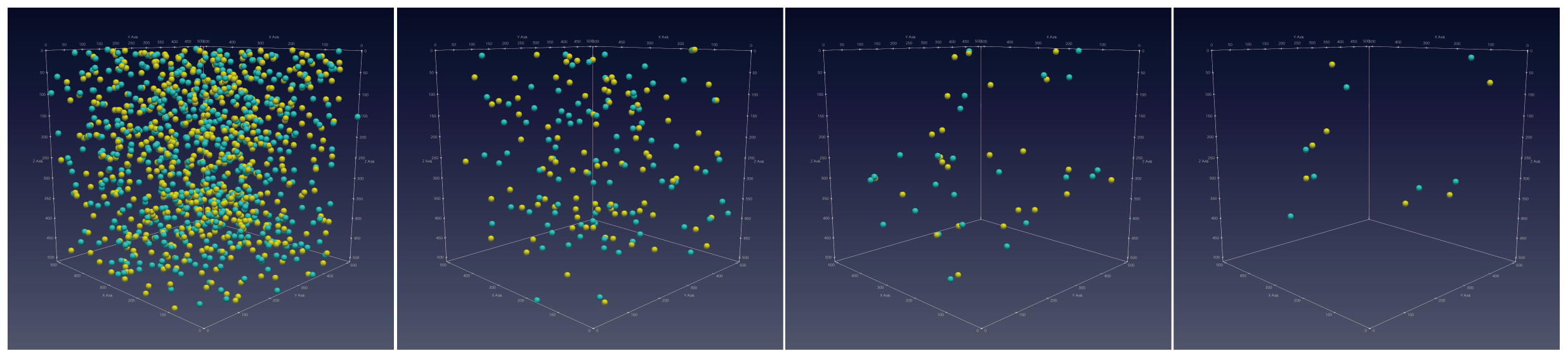}
\caption{
Snapshots of the evolution of the monopole network in the fat monopole regime from the simulation with $N^3 = 512^3$ grid points. Yellow and cyan points correspond to monopoles and anti-monopoles respectively. The corresponding simulation times are $v\tau = 1, 41, 81,$ and $121$, from left to right.
}
\label{fig:snapshot}
\end{figure*}

\begin{figure*}[tp]
\centering
\includegraphics[width = 8.5cm, clip]{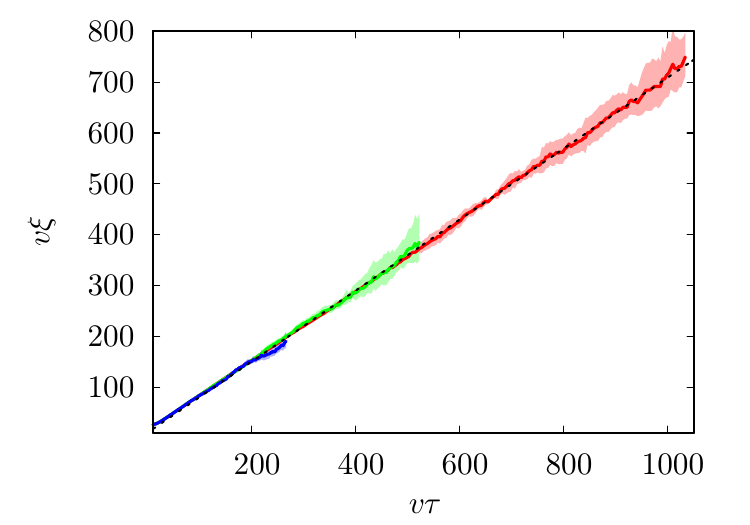}
\includegraphics[width = 8.5cm, clip]{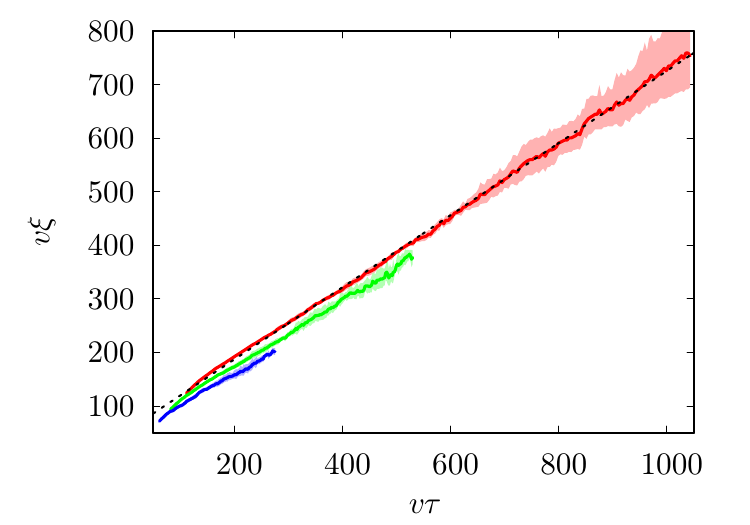}
\caption{
The evolution of the mean monopole separation in the fat (left) and physical (right) monopole cases. The blue, green and red colors correspond respectively to $N^3 = 1024^3$, $2048^3$ and $4096^3$. Solid lines and shaded regions correspond respectively to the average and 1 $\sigma$ error from 5 simulations with different initial data set.
The dashed lines are linear fitting from 4K simulations (see Table \ref{tab:scaling}). 
}
\label{fig:scaling}
\end{figure*}

\begin{table*}[tp]
    \centering
    \begin{tabular}{c c c c}
    \hline
    $N^3$ & $1024^3$ (1K) & $2048^3$ (2K) & $4096^3$ (4K) \\
    \hline
    fat & ~$(0.620 \pm 0.00693,~23.9 \pm 1.31)$~  & ~$(0.690 \pm 0.00290,~15.1 \pm 0.967)$~ & ~~$(0.698 \pm 0.000904,~10.9 \pm 0.566)$~ \\
    physical & ~$(0.564 \pm 0.00387,~46.3 \pm 0.756)$~ & ~$(0.614 \pm 0.00167,~50.9 \pm 0.567)$~ & ~$(0.674 \pm 0.00108,~52.3 \pm 0.692)$~ \\
    \hline
    \end{tabular}
    \caption{Linear fitting parameters $(A_{\rm sc},B_{\rm sc})$ for the mean monopole separation $v\xi = A_{\rm sc} v\tau + B_{\rm sc}$. We use the data with $v\tau > 100$ from 5 simulations.}
    \label{tab:scaling}
\end{table*}

In order to follow the nonlinear dynamics of the global monopole network, we employ numerical lattice simulations to directly solve the classical field equations for scalar fields (\ref{eq:KGeq}).
In our simulations, we adopt the conformal time as a simulation time variable and set the grid number $N^3 = 1024^3$, $2048^3$, $4096^3$ (which we call 1K, 2K, 4K respectively), the boxsize $L = (N/2)v^{-1}$, the initial conformal time $\tau_i = 10v^{-1}$, the initial scale factor $a_i = 1$.
We set the initial values following \cite{Hindmarsh:2025vxh}. 
First, we generate Gaussian random fields in the Fourier space with the power spectrum for the scalar field
\begin{align}
    P_{\rm ini}(k) = \frac{(k\ell_i)^3}{\sqrt{2\pi}}e^{-\frac{1}{2}(k \ell_i)^2},
\end{align}
where $\ell_i$ is the initial correlation length which we take $\ell_i = 10 v^{-1}$ in our simulations. Then, we perform the inverse Fourier transformation to get the field at each point in the real space.
However, the resultant field configuration still has a large artificial gradient energy. To remove it, we evolve the fields by the following diffusion equations,
\begin{align}
    \dot{\phi_i} - \nabla^2 \phi_i + \frac{\partial V}{\partial \phi_i} = 0,
\end{align}
which is called the initial cooling process \cite{Correia:2020gkj}. We set the duration of this process as $\tau_{\rm dif} = 10v^{-1}$ and update the fields using the first order Euler method with the stepsize $\Delta\tau = 0.15(L/N)^2$. Note that the cosmic expansion is turned off in this phase.
Then, we update the field 
using the second order Leap-frog method with the stepsize $\Delta\tau = 0.2(L/N)$ (see, e.g.,\cite{Figueroa:2020rrl,Baeza-Ballesteros:2025tme} for detailed formulation of the cosmological lattice simulation).

To justify the scaling evolution of global monopoles, we need to follow the dynamics of the network for a long time. However, as the Universe expands, the monopole core size becomes smaller and smaller compared with the simulation box, which means that we need more resolution at later time. To avoid it, one can adopt the so-called fat monopole approximation (or PRS formulation) where the monopole core size increases linearly with the scale factor \cite{Press:1989yh}. Practically, we replace the constant $\lambda$ with the time-dependent one, $\lambda(t) = \lambda_0 a^{2(s-1)}$. One can recover the realistic or ``physical'' situation for $s=1$ but we set $s=0$ for the fat monopole approximation. In what follows, we study both the fat and the physical monopole regimes.

In the main evolution phase, we evolve the fields for a half light crossing time, $\tau_{\rm LC}/2 = L/2$.
However, in the physical monopole case, we insert the so-called core-growth or extra-fattening phase, where we set $\lambda(t) = \lambda_0/a^4$, i.e. $s = -1$, before the main evolution phase.
After this phase, the monopole size is enlarged, which enables us to follow the dynamics for a half light crossing time in the main evolution phase. The simulation ends when the monopole size becomes the same as the initial one before the core-growth phase.

To identify the monopole, we compute the local winding number in each cubic cell as formulated in \cite{Antunes:2002ss}.
Fig.~\ref{fig:snapshot} shows snapshots of the evolution of the monopole network. This figure illustrates that the number of monopoles decreases with time due to the monopole-antimonopole pair annihilation.

\subsection{Scaling law}

The scaling evolution is a characteristic feature of the global monopole network.
To see this, we count the monopole number in the simulation box, $N_{\rm mn}$, using the local winding number and compute the mean comoving distance of the monopole separation,
\begin{align}
    \xi = \left( \frac{V}{N_{\rm mn}} \right)^{1/3},
\end{align}
where $V = L^3$ is the comoving volume of the simulation box.
Fig.~\ref{fig:scaling} shows the evolution of the mean (comoving) monopole separation using data from 5 simulations with different sets of random initial values.
The figure shows that the mean monopole separation grows linearly with conformal time, as in the case with the local cosmic string~\cite{Kibble:1984hp}. 
Indeed, the data can be well-fitted by the linear function, $v\xi = A_{\rm sc} v\tau + B_{\rm sc}$, and the fitting parameters $(A_{\rm sc},B_{\rm sc})$ using the data with $v\tau > 100$ are summarized in Table \ref{tab:scaling}.

\subsection{NG boson production}

\begin{figure*}[tp]
\centering
\includegraphics[width = 8.5cm, clip]{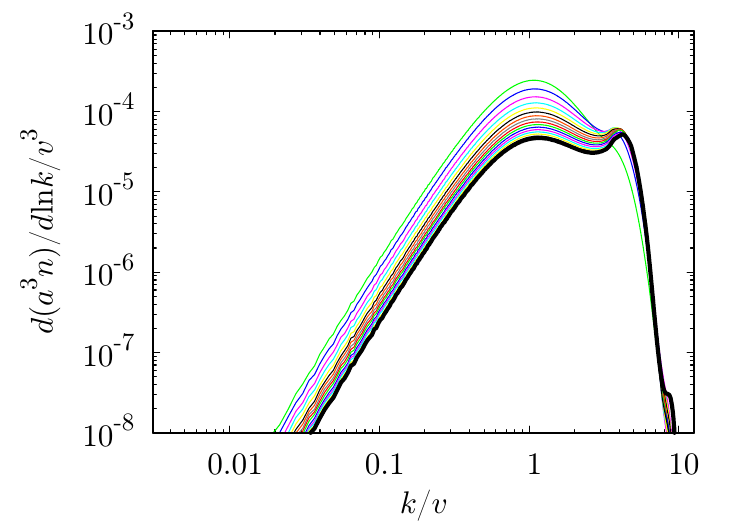}
\includegraphics[width = 8.5cm, clip]{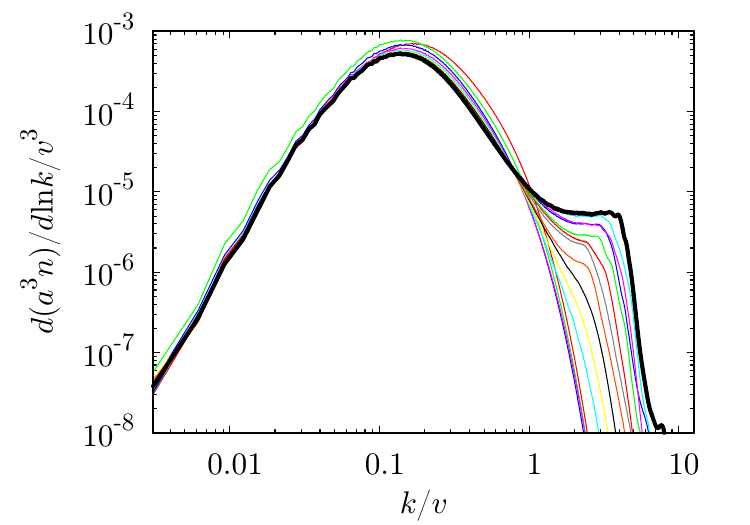}
\caption{
Spectrum of the comoving number density of the radial mode in the fat (left) and the physical (right) monopole regimes. The thick black line shows the final time, $\tau = 1034v^{-1}$ ($1043.6 v^{-1}$) for the fat (physical) monopole regime, and the time interval of each plot is $\Delta \tau = 64v^{-1}$.
}
\label{fig:spectrum_R}
\end{figure*}

\begin{figure*}[tp]
\centering
\includegraphics[width = 8.5cm, clip]{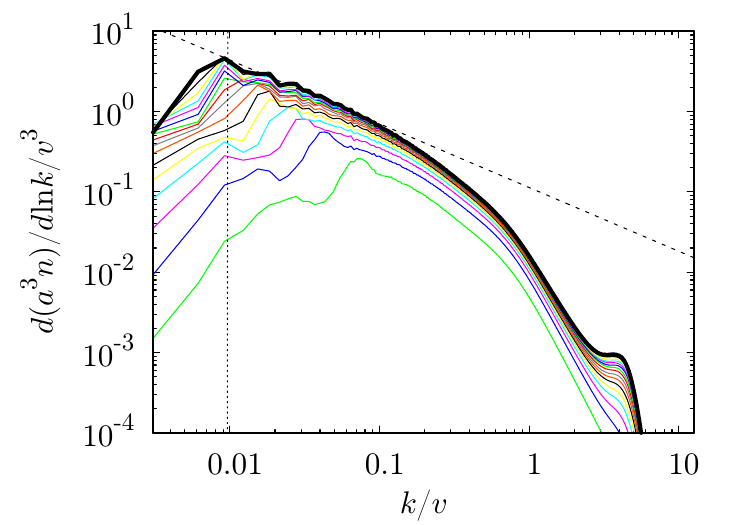}
\includegraphics[width = 8.5cm, clip]{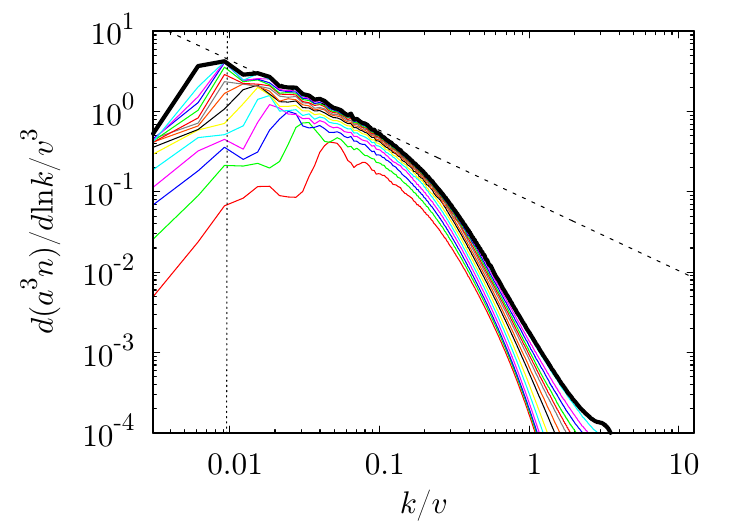}
\caption{
Spectrum of the comoving number density of the first NG mode in the fat (left) and the physical (right) monopole regimes. The thick black line shows the final time, $\tau = 1034v^{-1}$ ($1043.6 v^{-1}$) for the fat (physical) monopole regime, and the time interval of each plot is $\Delta \tau = 64v^{-1}$. The vertical dotted line corresponds to $k/a = 10H$ at the final time. The dashed line is the fitting of the final spectrum $A(k/v)^p$ with $(p,A) = (-0.795,0.113)$ (left), $(-0.875,0.779)$ (right) using the data with $0.01 < k/v < 0.1$.
}
\label{fig:spectrum_NG1}
\end{figure*}

\begin{figure*}[tp]
\centering
\includegraphics[width = 8.5cm, clip]{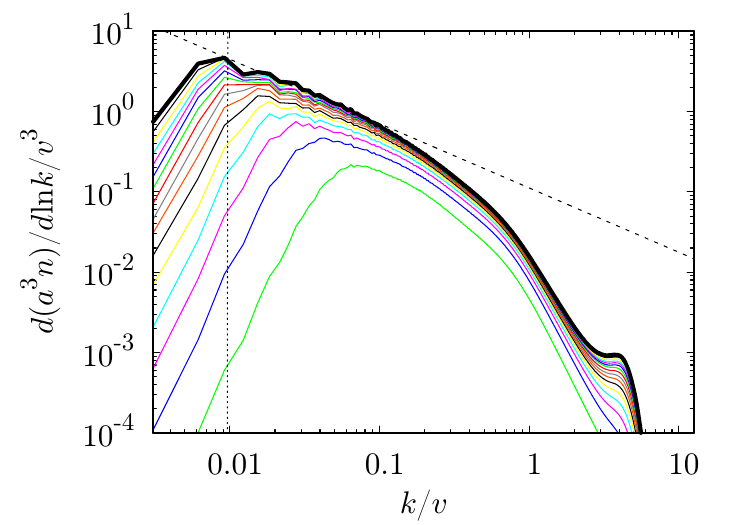}
\includegraphics[width = 8.5cm, clip]{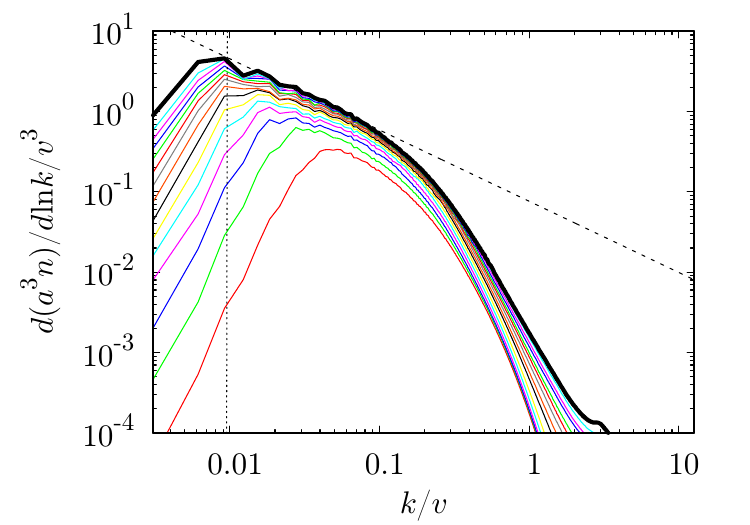}
\caption{
Spectrum of the comoving number density of the second NG mode in the fat (left) and the physical (right) monopole regimes. The thick black line shows the final time, $\tau = 1034v^{-1}$ ($1043.6 v^{-1}$) for the fat (physical) monopole regime, and the time interval of each plot is $\Delta \tau = 64v^{-1}$. The vertical dotted line corresponds to $k/a = 10H$ at the final time.
The dashed line is the fitting of the final spectrum $A(k/v)^p$ with $(p,A) = (-0.801,0.112)$ (left), $(-0.887,0.0761)$ (right) using the data with $0.01 < k/v < 0.1$.
}
\label{fig:spectrum_NG2}
\end{figure*}

To get the late time abundance of the emitted NG bosons, we need to compute the evolution of the comoving number density, which would be conserved in the expanding universe without particle emissions. 
Figs.~\ref{fig:spectrum_R}, \ref{fig:spectrum_NG1}, \ref{fig:spectrum_NG2} show the evolutions of the spectra of the comoving number density for the radial mode and two NG modes. They show that the production of the radial mode is suppressed compared with NG modes. The spectrum of each NG mode has an IR peak corresponding to the Hubble scale, as in the case with the global/semilocal string~\cite{Yamaguchi:1998gx,Kanda:2025hgi}. 
The spectrum around the peak can be fitted by the power law function $P(k) = A(k/v)^p$. Using the data with $0.01 < k/v < 0.1$, we obtained the power as $p = -0.795 \pm 0.0215$ (NG1) and $p = -0.801 \pm 0.0243$ (NG2) for the fat monopole case, and $p = -0.875 \pm 0.0246$ (NG1) and $p = -0.887 \pm 0.0266$ (NG2) for the physical monopole case.
These spectra for NG modes are similar to those in the semilocal string case \cite{Kanda:2025hgi}.

\begin{figure*}[tp]
\centering
\includegraphics[width = 8.5cm, clip]{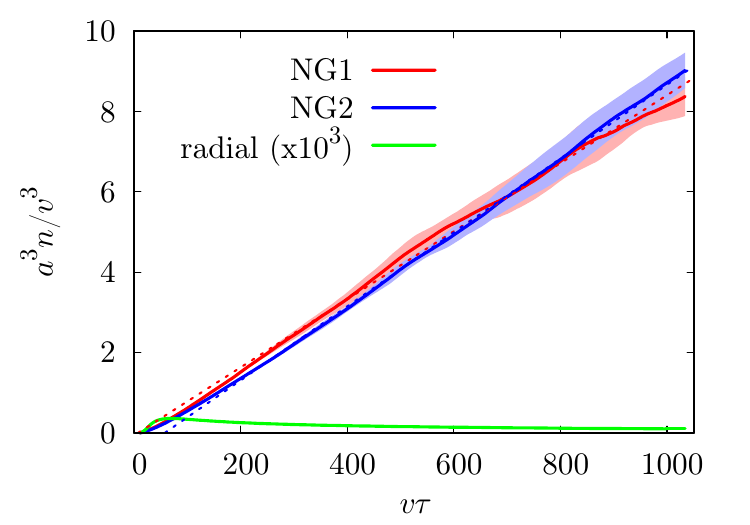}
\includegraphics[width = 8.5cm, clip]{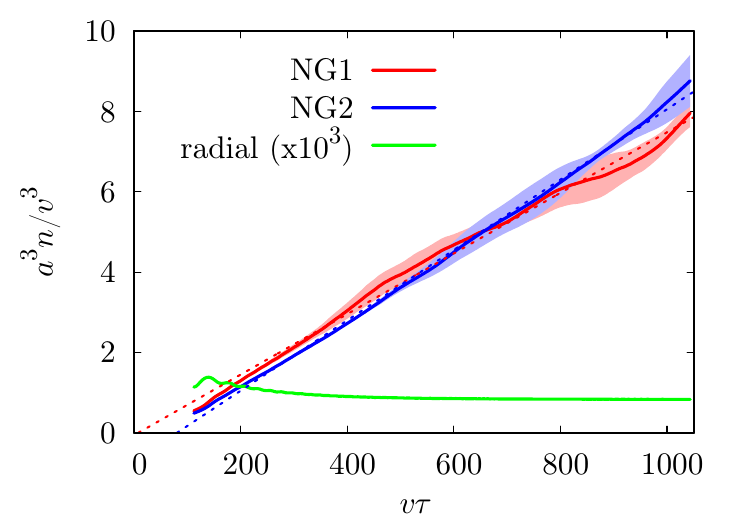}
\caption{
The evolution of the mean comoving number density of the two NG modes (red and blue) and radial mode (green) in the fat (left) and physical (right) monopole cases from the simulations with $N^3 = 4096^3$.
Solid lines and shaded regions correspond respectively to the average and 1 $\sigma$ error from 5 simulations with different initial data set.
The dashed lines are linear fitting (see Table \ref{tab:nNG}).
}
\label{fig:evolve_n}
\end{figure*}

From the spectra one can obtain the evolution of the comoving number density of the emitted particles. Fig.~\ref{fig:evolve_n} shows the evolution of the radial mode (green), NG1 and NG2 modes (red and blue).
The emission of the radial mode is suppressed, but the comoving number density of each NG mode increases linearly with the conformal time as fitted by the dotted line.
The linear fitting parameters are summarized in Table \ref{tab:nNG}.
Then, the evolution of the number density of each NG mode can be expressed as $n_{\rm NG} = C (v\tau_i)^2 v^2 H$ with $C$ the numerical coefficient measured from the numerical data. 
Specifically, we obtain $n_{NG} \simeq 0.7$\,-\,$0.9 v^2 H$ from the linear fitting. 
This shows a good agreement with the case with the NG boson emission from the semilocal string system \cite{Kanda:2025hgi}.

\begin{table*}[tp]
    \centering
    \begin{tabular}{c c c c}
    \hline
    & NG1 & NG2 \\
    \hline
    fat & ~$(0.00847 \pm 3.17 \times10^{-5},~-0.0602 \pm 0.0199)$~  & ~$(0.00920 \pm 1.31 \times 10^{-5},~-0.528 \pm 0.00820)$~ \\
    physical & ~$(0.00754 \pm 3.99 \times 10^{-5} ,~-0.0547 \pm 0.0255)$~ & ~$(0.00876 \pm 2.30 \times 10^{-5},~-0.700 \pm 0.0147)$~ \\
    \hline
    \end{tabular}
    \caption{Linear fitting parameters $(A_{\rm NG},B_{\rm NG})$ for the comoving number density of NG modes $a^3 n_{\rm NG}/v^3 = A_{\rm NG} v\tau + B_{\rm NG}$.
    We use the data with $v\tau > 100$ from 5 realizations of 4K simulation.
    }
    \label{tab:nNG}
\end{table*}

The abundance of emitted particles is related to the scaling evolution of the global monopole. 
Denoting the mean physical distance of the monopole separation by $\xi_{\rm phys} \equiv a\xi$, we can estimate the effective monopole mass as $m_M \simeq 4\pi v^2\xi_{\rm phys}$. The energy density of the monopole network is therefore given by
\begin{align}
    \rho_M \simeq \frac{m_M}{\xi_{\rm phys}^3} = \frac{4\pi v^2}{\xi_{\rm phys}^2} .
\end{align}
In the scaling regime, the monopole energy density scales as $\rho_M\propto t^{-2}$.
By contrast, if the monopoles were to remain approximately fixed in comoving coordinates, their mean physical separation would scale as $\xi_{\rm phys}^{\rm free}\propto a$, and hence the network energy density would scale as $\rho_M^{\rm free} \propto t^{-1}$. 
Normalizing the two energy densities at an arbitrary reference time $t_0$, such that $\rho_M(t_0)=\rho_M^{\rm free}(t_0)$, we estimate the instantaneous energy loss rate per unit physical volume as
\begin{align}
    \Gamma(t) &= \left[\dot{\rho}_M^{\rm free}(t_0) - \dot{\rho}_M(t_0) \right]_{t_0\to t} \nonumber\\
    &= \left[2\frac{\dot{\xi}_{\rm phys}(t)}{\xi_{\rm phys}(t)} - \frac{1}{t}\right] \rho_M
    \xrightarrow{\tau\gg70v^{-1}} 2H\rho_M.
\end{align}

If the energy loss of the monopole network is entirely transferred to the emitted NG bosons, their production can be treated in a manner analogous to that discussed in Sec.~4 of Ref.~\cite{Gorghetto:2018myk}. 
For a soft emission spectrum of the form $\frac{\partial\Gamma}{\partial k} \propto k^{-q}$, where $q>1$, the number density of the emitted NG bosons is estimated as
\begin{align}
    n_{\rm NG} \simeq \frac{16\pi t^2 C_{\rm spec}}{\xi_{\rm phys}^2} v^2H,
    \label{Eq:n_NG_analys}
\end{align}
where $C_{\rm spec}$ is a dimensionless numerical factor determined by the emission spectrum.\footnote{Note that the spectral index $q$ of $\frac{\partial\Gamma}{\partial k} \propto k^{-q}$ should not be confused with the index $p$ obtained by fitting the spectra shown in FIG.~\ref{fig:spectrum_NG1} and \ref{fig:spectrum_NG2}. $q$ describes an effective spectrum for which the UV cutoff is set by the radial-mode mass, while the IR one scales linearly with $H$. See Ref.~\cite{Gorghetto:2018myk} for further details.}
Since $\xi_{\rm phys}\propto t$, Eq.~\eqref{Eq:n_NG_analys} reproduces the linear dependence $n_{\rm NG}\propto H$ observed in our simulations. By fitting this expression to our simulations, we find $C_{\rm spec}\simeq0.03$.

\section{PNGB DM abundance} \label{sec:pNGDM}

The emitted NG bosons may contribute to the dark matter abundance if either or both of the NG modes acquires a mass during the subsequent evolutions.
This possibility arises in a variety of well-motivated particle-physics scenarios, including axion models. 
The mass can also be generated by introducing a soft symmetry-breaking term, such as $m_{\rm NG}^2 \phi_i^2/2$, into the scalar potential.
This possibility motivated us to investigate whether the emitted pseudo NG (pNG) bosons can account for a significant fraction of the observed dark matter abundance.

We now estimate the relic abundance of the emitted pNG bosons. 
Assuming that the emissions of such massive pNG bosons stop when the Hubble parameter, which is the typical energy of emitted NG bosons, becomes smaller than the mass, the abundance of such pNG DM is fixed at that time. The relic abundance of the pNG DM can be estimated as follows \cite{Kanda:2025hgi},
\begin{align}
\Omega_{\rm NG}h^2 &= \frac{m_{\rm NG} n_{{\rm NG},0}/s_0}{\rho_{{\rm cr},0}/s_0 h^{-2}} \nonumber \\
    &\simeq 0.2 F(m_r/m_{\rm NG}) \left( \frac{m_{\rm NG}}{10^{-13} {\rm eV}} \right)^{1/2} \left( \frac{v}{10^{14} {\rm GeV}} \right)^2.
\end{align}
where $F(m_r/m_{\rm NG})$ is a factor parameterizing possible deviations from the exact scaling. We define as
\begin{align}
    F(m_r/H) \equiv \left(\frac{\zeta(m_r/H)}{\zeta_{\rm sc}}\right)^{2/3},
\end{align}
where $\zeta(m_r/H)\equiv(\xi_{\rm phys} H)^{-3}$ is the average number of monopoles per Hubble volume, and $\zeta_{\rm sc}$ denotes its constant value in the exact scaling regime. If
\begin{align}
    \zeta(m_r/H) = A_{\rm log} + B_{\rm log} \log(m_r/H),
\end{align}
as suggested in Ref.~\cite{Nakano:2026zme}, $F(m_r/m_{\rm NG})$ gives a enhancement over the exact scaling result.

\begin{figure}[tp]
\centering
\includegraphics [width = 8.5cm, clip]{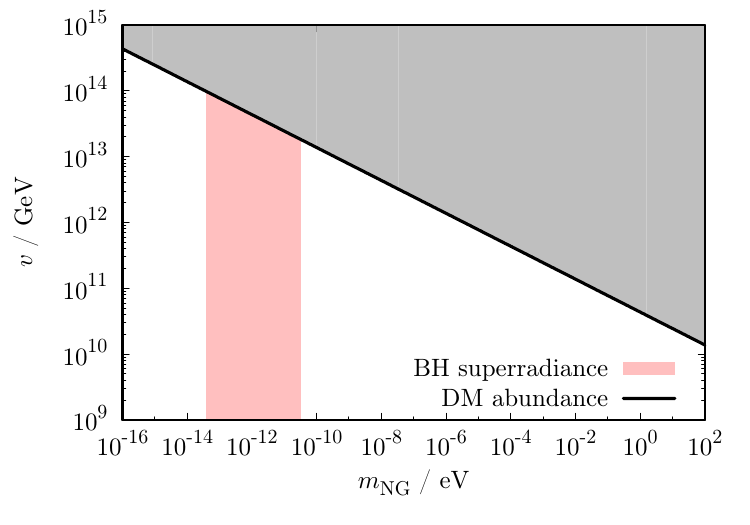}
\caption{
Allowed parameter region in $m_{\rm NG}$-$v$ plane. The current relic abundance of dark matter can be explained on the thick black line, i.e., $\Omega_{\rm NG} = \Omega_{\rm DM}$.
The gray and red shaded regions are ruled out due to the overabundance of dark matter and the black hole superradiance~\cite{Cardoso:2018tly}, respectively.
}
\label{fig:constraint}
\end{figure}

Fig.~\ref{fig:constraint} shows an allowed parameter region for the scaling monopole network we observed. The current DM abundance can be explained on the thick black line. Thus, pNG DM can be realized in a wide mass range, which can be as small as $m_{\rm NG} \sim 10^{-10}$ eV.

\section{Discussion} \label{sec:discussion}

In this paper, we have shown the emission of the NG modes from the scaling evolution of global monopoles by using large-scale numerical lattice simulations. Our data show that the mean monopole separation grows linearly with time, which verifies the scaling evolution. 
The spectrum of emitted NG modes has an IR peak around the Hubble scale, while the emission of the radial mode is suppressed.
Moreover, we have shown that the number density scales linearly with the Hubble scale. Combined with the scaling behavior of the network, this implies that the free NG radiation is dominated by soft modes.

The shape of the spectrum and the evolution of the comoving number density of the emitted NG bosons are remarkably similar to those observed in semilocal string networks~\cite{Kanda:2025hgi}. This similarity strongly suggests that the endpoint dynamics, rather than the dynamics of string interiors, is the dominant source of NG boson emissions from semilocal string networks. Although the spectral similarity alone does not directly identify the main emission region, our results support the global monopole picture of semilocal string endpoints and provide a consistent physical interpretation of the efficient NG boson production observed in our previous simulations. Our results thus indicate that the NG boson emission from global monopoles, and semilocal string networks can be understood within a unified physical picture. We will develop an analytical model of this emission mechanism in the subsequent paper.

Although our data show consistency with the scaling evolution, the logarithmic scaling violation is reported in \cite{Nakano:2026zme} as in the case with the global string \cite{Gorghetto:2018myk,Kawasaki:2018bzv,Gorghetto:2020qws,Buschmann:2021sdq,Saikawa:2024bta,Benabou:2024msj}.
To verify the scaling violation, we need longer simulations with various initial conditions. It is left for future work.

Gravitational waves can be emitted from the monopole dynamics itself \cite{Aburatani:2026rct}. However, the constraint is not significant in the parameter range of our interest, e.g., $v \lesssim 10^{16}$ GeV or $m_{\rm NG} > 10^{-20}$ eV. Similarly, the constraint from the PBH abundance \cite{Aburatani:2026rct} is also irrelevant in this case.

Inclusion of the soft mass term is not trivial in the current case, since it can significantly modify the vacuum structure.
Namely, the global strings appear if one of the NG modes has a mass and domain walls are formed if both NG modes have masses. In the former case, the massless NG bosons are still continuously emitted, but their abundance is negligible compared to that of the pNG mode. Also, gravitational waves can be emitted from such global strings \cite{Chang:2019mza,Gorghetto:2021fsn,Chang:2021afa}. In the latter case, the domain walls may dominate the universe and it is problematic. Such domain walls should be annihilated before they dominate the universe. It can be achieved if, e.g., we introduce a bias in the potential. Detailed study of such multi-stage defect formation and its cosmological implications, e.g., particle radiation and gravitational waves, is left for future work.

\section*{Acknowledgments}
This work is supported by JSPS KAKENHI Grant No. JP26K17132 (Y.K.).
This work used computational resources of Fugaku supercomputer, provided by RIKEN Center for Computational Sciences, through the HPCI System Research Project (Project ID: hp250177, hp260117). 
The authors thank Keisuke Harigaya and Fuminobu Takahashi for useful comments. 




\bibliography{ref.bib}

\end{document}